\documentclass[f]{ceurart}
\usepackage{graphicx} 
\usepackage{mwe}
\usepackage{wrapfig}

\usepackage[utf8]{inputenc}
\usepackage{acronym}
\usepackage{multirow}
\usepackage{todonotes}
\usepackage{soul}
\usepackage{hyperref}
\usepackage[ruled,vlined]{algorithm2e}
\usepackage{listings}
\usepackage{array}
\usepackage{tabularx}
\usepackage[inkscapeformat=png]{svg}
\usepackage{amsmath,amssymb,amsfonts}
\usepackage{pifont}
\usepackage[normalem]{ulem}
\usepackage{flushend}

\usepackage{booktabs}

\title{An Empirical Study of Gas Optimization Practices in Solidity Smart Contracts}
\date{March 2026}

\begin{document}

\copyrightyear{2024}
\copyrightclause{Copyright for this paper by its authors.
  Use permitted under Creative Commons License Attribution 4.0
  International (CC BY 4.0).}

\author[1]{Francesco Salzano}[%
orcid=0000-0002-1029-4861,
email=francesco.salzano@unimol.it
]
\cormark[1]
\address[1]{University of Molise,
  StakeLab, Italy}

\author[2]{Paolo Martino}[%
email=p.martino3@studenti.unimol.it
]

\author[3]{Simone Scalabrino}[%
orcid=0000-0003-1764-9685,
email=simone.scalabrino@unimol.it
]

\author[4]{Rocco Oliveto}[%
orcid=0000-0002-7995-8582,
email=rocco.oliveto@unimol.it
]

\author[5]{Remo Pareschi}[%
orcid=0000-0002-4912-582X,
email=remo.pareschi@unimol.it
]
\cortext[1]{Corresponding author.}

\begin{abstract}
In the Ethereum ecosystem, smart contract execution is governed by a gas model where every computational operation incurs a direct economic cost. Consequently, gas efficiency is a critical requirement for the economic sustainability and widespread adoption of decentralized applications. Despite the importance of cost-effective development, there remains a significant gap between theoretical optimization patterns and the actual practices applied by developers in real-world projects. 
This paper presents an empirical study investigating gas optimization practices in open-source Solidity repositories created on GitHub between 2018 and 2025. Adopting a Mining Software Repositories approach, we analyzed a dataset of gas-related commits identified through a multi-stage pipeline involving NLP-based filtering and manual qualitative inspection. Based on the analysis of 400 representative commits, we derive an empirical taxonomy of gas optimization strategies consisting of 8 macro-categories and 36 subcategories. 
Our findings reveal that optimization efforts span multiple levels of abstraction, from low-level arithmetic refinements and bytecode-size reductions to broader architectural changes such as logic internalization and external call reduction. Furthermore, we observe that developers often manage complex trade-offs between execution frequency, safety checks, and computational costs, highlighting that effective optimization is inherently context-dependent. This study provides a practical knowledge base for the design of future developer support tools and advances the understanding of cost-aware software engineering in the blockchain domain.
\end{abstract}


\conference{DLT2026: 8th Distributed Ledger Technologies Workshop, Pula, Italy, June, 4-6 2026}

\maketitle

\section{Introduction}

The emergence of blockchain technology and the widespread adoption of Ethereum have introduced a major shift in software engineering. By enabling decentralized programs to execute on a distributed infrastructure, smart contracts make it possible to manage digital assets and application logic without relying on centralized intermediaries \cite{zheng2017overview, wood2014ethereum}. This setting, however, introduces a constraint that is largely absent in traditional software systems: execution has a direct economic cost.

In Ethereum, every computational step performed by a smart contract consumes gas, the unit used to price the execution of operations in the Ethereum Virtual Machine \cite{wood2014ethereum}. As a consequence, code efficiency is not merely a performance concern, but a core requirement for economic sustainability. Contracts that are not carefully optimized may impose excessive fees on users, reduce the practicality of decentralized applications, and ultimately hinder adoption \cite{marchesi2020design}.

Despite the importance of gas efficiency, the current understanding of how developers actually optimize smart contracts remains limited. Existing research has introduced static analysis tools and optimization guidelines aimed at detecting inefficient patterns or suggesting improvements \cite{albert2020gasol}. Yet prior studies indicate that automatic recommendations alone are often insufficient to capture the complexity of real development decisions, and that developers frequently rely on manual interventions grounded in practical experience and knowledge of the Ethereum execution model \cite{brandstatter2020characterizing}.

This creates a gap between optimization knowledge formalized in the literature and optimization practices applied in real projects. Many gas saving techniques circulate through developer discussions, technical blogs, and informal documentation, while only a limited portion has been systematically studied through empirical evidence \cite{chen2017under}. This paper addresses that gap by investigating how developers of open source Solidity projects concretely optimize gas consumption in practice.

To answer this question, analyzing only the final version of smart contracts is not sufficient. It is necessary to observe how code evolves over time and how developers explicitly introduce cost-oriented changes. For this reason, we adopt a Mining Software Repositories approach and analyze the history of Solidity repositories hosted on GitHub. More specifically, our study focuses on commits, since they provide a fine-grained view of code evolution together with the intent expressed in commit messages. By comparing the code before and after modifications associated with gas optimization, we identify the concrete strategies used to reduce execution costs.

Through a multi-stage pipeline combining automated filtering and manual qualitative inspection, we construct a dataset of gas-related commits drawn from heterogeneous Solidity repositories created between 2018 and 2025. We then analyze these changes through open coding in order to derive a taxonomy grounded in observed development practice.

This paper makes the following contributions. First, it provides an empirical characterization of gas optimization activity in open source Solidity development through the analysis of commit level changes. Second, it derives a taxonomy of optimization strategies observed in practice, consisting of \texttt{8} macro categories and \texttt{36} subcategories. Third, it highlights that gas optimization spans multiple levels of intervention, ranging from localized code changes to broader structural refactorings, thereby offering a practical knowledge base for future research on smart contract analysis and developer support tools.
All the data and the code used in our analysis are available on our replication package\cite{salzano2026gasoptimizationreplication}.

The remainder of the paper is organized as follows. Section~\ref{sec:background} introduces the necessary background on smart contracts, Solidity, Ethereum, and the gas model. Section~\ref{sec:related} discusses the most relevant literature. Section~\ref{sec:design} presents the empirical study design. Section~\ref{sec:results} reports the resulting taxonomy and the main findings. Section~\ref{sec:discussion} discusses the implications of the study, while Section~\ref{sec:treats} presents the threats to validity of the study and its limitations. Finally, Section~\ref{sec:conclusion} concludes the paper.

\section{Background}
\label{sec:background}
This section provides the technical background needed to understand the study, covering smart contracts, Solidity, Ethereum execution, and the gas model.

\subsection{Smart Contracts and Solidity}

Smart contracts are event-driven programs deployed on a blockchain and replicated across multiple nodes in a decentralized network. They automatically execute predefined logic when specific conditions are satisfied, enabling trustless interactions without relying on centralized intermediaries~\cite{zou2019smart}. Once deployed, a smart contract is uniquely identified by its address, and its code becomes immutable. Although immutability is a core property, update mechanisms such as proxy patterns can be adopted to redirect calls to newer implementations while preserving the original contract state and address~\cite{bodell2023proxy}.

Interactions with smart contracts occur through transactions issued either by external users or by other contracts. Each transaction triggers the execution of the contract logic on all participating nodes, which must reach the same result to maintain consensus. The outcome of execution is then recorded on the blockchain, making it irreversible due to the append-only nature of the ledger. As a consequence, unexpected failures during execution may lead to outcomes that cannot be undone, highlighting the importance of writing reliable and efficient code.

Solidity is the most widely used high level programming language for implementing smart contracts on Ethereum. It provides abstractions similar to object oriented languages while compiling into low level bytecode executed by the underlying virtual machine. Developers use Solidity to define contract state, functions, and interactions with other contracts, ultimately shaping both the functional behavior and the resource consumption of decentralized applications.

\subsection{Ethereum and Gas Model}

Ethereum is the most widely adopted platform for smart contract execution and introduced a general purpose computational model through the Ethereum Virtual Machine, an execution environment that runs contract bytecode consistently across all nodes~\cite{wood2014ethereum,wohrer2020domain}. In this setting, every node independently executes the same instructions, ensuring deterministic outcomes and agreement on the global state.

Executing arbitrary code in a decentralized environment raises challenges related to resource management. To address this issue, Ethereum adopts a cost model based on gas, which represents the unit used to measure the computational effort required to perform operations~\cite{zou2019smart}. Each low level instruction executed by the virtual machine is associated with a predefined gas cost. Simple operations such as arithmetic computations require minimal gas, while more expensive actions, such as writing data to persistent storage, incur significantly higher costs.

Every transaction specifies a gas limit, which defines the maximum amount of computational effort that the sender is willing to pay for. During execution, gas is consumed incrementally as instructions are processed. If execution exceeds the specified limit, the transaction is reverted and all state changes are discarded, while the gas spent up to that point is not refunded~\cite{chen2020defining}. This mechanism prevents infinite execution and ensures fair allocation of network resources.

Within this economic model, efficiency becomes a fundamental requirement. Since users pay for every executed operation, inefficient implementations directly translate into higher costs. As a result, optimizing gas consumption is essential for improving usability and making decentralized applications economically sustainable.

\section{Related Work}
\label{sec:related}
This section presents a review of prior research on smart contract maintenance and gas-related analyses, and positions our study within the context of the existing body of literature.

Research on Ethereum smart contracts has repeatedly shown that gas consumption is a central dimension of software quality, since execution costs directly affect both usability and adoption. Prior work has addressed this issue from several complementary angles, including smart contract maintenance, detection of gas-related inefficiencies, repository mining methodologies, design-oriented optimization knowledge, and automated analysis tools. These studies provide the conceptual and methodological foundation for our investigation, while also highlighting the lack of empirical evidence on how developers perform gas optimization in real-world Solidity projects.

A first relevant perspective comes from the study of smart contract maintenance and evolution. Rosa et al. \cite{rosa2025and} investigated why and how developers maintain Solidity smart contracts, classifying 590 commits according to established maintenance categories such as corrective, perfective, adaptive, and preventive maintenance. Their results offer an important view of the broader evolution of smart contract software, but they also suggest that gas optimization is only marginally visible within generic maintenance analyses. In their qualitative sample, only a limited number of commits were explicitly associated with gas-related improvements, and these changes were often intertwined with broader refactorings. This observation is especially relevant for our work, since it indicates that gas optimization is difficult to isolate without a dedicated research design. Our study builds on this insight by focusing exclusively on commits that target gas reduction, with the goal of characterizing the concrete techniques adopted by developers.

A second line of work has examined the economic consequences of implementation decisions in Solidity. Di Sorbo et al. \cite{di2022profiling} showed that specific coding choices, framed as Solidity code smells, can significantly affect gas consumption. Their study provides quantitative evidence that inefficient implementations may impose substantial economic costs on users, reinforcing the idea that optimization is not merely a stylistic concern. At the same time, this body of work mainly concentrates on the identification and measurement of inefficiencies. Our contribution takes a different perspective: instead of detecting problematic patterns in deployed code, we study how developers modify contracts over time in order to address these costs. In this sense, our work complements gas profiling research by moving from inefficiency detection to the empirical observation of optimization practices.

From a methodological standpoint, our study is closely related to recent work applying Mining Software Repositories techniques to blockchain software. Salzano et al. \cite{salzano2026bridging}, in their comparative study on smart contract vulnerabilities, defined a rigorous protocol that combines Natural Language Processing-based commit filtering with manual qualitative inspection. Their approach demonstrates the value of mining repository histories to uncover development practices that are not directly visible from the final state of the code. We adopt a similar strategy in the gas domain, using commit message processing, lemmatization, and semantic filtering to identify candidate gas-related modifications, followed by manual validation and classification. The methodological affinity with this line of work strengthens the reliability of our dataset construction process and confirms the suitability of repository mining for studying smart contract engineering practices.

Another important reference is the literature on design knowledge for gas-efficient smart contract development. Marchesi et al. \cite{9050163} proposed one of the earliest structured collections of gas optimization design patterns for Ethereum, including practices such as variable packing and loop optimization. Their contribution is particularly relevant because it formalizes optimization strategies at the level of programming patterns and development guidelines. Yet these patterns are primarily derived from technical reasoning and controlled observations rather than from large-scale evidence of how developers actually revise code in open source repositories. Our work connects to this literature by examining whether the techniques observed in practice align with, extend, or diverge from the optimization strategies previously described in the academic literature. This empirical perspective allows us to assess how theoretical recommendations are reflected in real development workflows.

In parallel, several studies have introduced automated tools for gas analysis and optimization. GASOL applies static analysis to estimate upper bounds on execution costs and to identify opportunities for optimization \cite{albert2020gasol}, particularly with respect to state variables and expensive computations. GasChecker scales this idea to large collections of smart contracts \cite{chen2020gaschecker}, automatically detecting inefficient patterns such as redundant checks and unused variables with high precision. Other work has moved beyond detection and explored automatic code transformation. The refinement approach proposed by Nelaturu et al. aims to rewrite inefficient contract fragments \cite{nelaturu2021smart}, especially loops, in ways that reduce gas consumption while preserving semantics. These contributions show the growing maturity of automation in this area, but they also tend to focus on localized inefficiencies that can be recognized or transformed automatically. Our study addresses a different question, namely how developers themselves restructure contracts in practice, including changes that may involve contextual reasoning, multiple simultaneous interventions, or broader architectural revisions that are difficult to capture with static tools alone.

Gas efficiency has also been discussed in relation to security. \textit{eTainter} \cite{ghaleb2022etainter} shows that costly execution paths can be exploited to mount denial of service attacks, since adversaries may craft inputs that force contracts into overly expensive executions and exhaust available gas. This connection is important because it frames optimization not only as a matter of economic performance but also as a factor that can affect robustness and service availability. The existence of this link further motivates the need for a systematic understanding of gas-related development practices.

Taken together, prior research has established that gas matters, that implementation choices can strongly influence execution costs, and that tools and design recommendations can help detect or prevent inefficient patterns. What remains less understood is how developers concretely act on these concerns during the evolution of real smart contracts. Existing studies either focus on maintenance at a broad level, on the measurement of inefficiencies, on theoretical optimization patterns, or on automated detection and rewriting techniques. Our work fills this gap through an empirical analysis of gas-related commits in Solidity repositories, aiming to derive a taxonomy grounded in observed development activity rather than predefined recommendations.

\section{Empirical Study Design}
\label{sec:design}
In this section, we describe the design of our empirical study, including repository selection, commit mining, NLP-based filtering, sampling, manual inspection, and the taxonomy construction approach.

The purpose of this study is to characterize how developers optimize gas consumption in Solidity smart contracts, and to derive an empirical taxonomy of the optimization strategies adopted in practice. The study is aimed at researchers and practitioners interested in smart contract engineering, software quality, and cost-aware development on Ethereum. The context of the study is a dataset of gas-related commits mined from public Solidity repositories hosted on GitHub.

To achieve this goal, we analyzed a large set of commits from open source Solidity projects and followed a linear pipeline structured into five main phases: repository collection, commit mining and NLP-based filtering, dataset refinement, stratified sampling, and manual inspection with open coding. This design allows us to combine large-scale automated mining with qualitative analysis of concrete code changes.

The study is guided by the following research question:

\begin{itemize}
    \item \textbf{RQ$_1$}: How do developers of open source Solidity projects optimize the gas consumption of smart contracts?
\end{itemize}

\subsection{Data Collection}
\label{sec:context}

The context of our study is a dataset of commits potentially related to gas optimization in Solidity smart contracts. These commits provide pairs of pre-change and post-change code, making it possible to observe how developers revise contracts to reduce execution costs.

\subsubsection{Repository Selection and Commit Mining}

We focused our study on Solidity smart contracts by collecting repositories from GitHub. To improve the quality and relevance of the dataset, we applied a set of filters based on repository metadata.

First, we included only repositories containing Solidity code and having at least 10 stars. The number of stars is commonly used as a proxy for popularity and perceived usefulness in the GitHub ecosystem, and has been adopted in prior software engineering studies for repository selection~\cite{dabic2021sampling,rosa2023quality,salzano2026bridging}. Restricting the dataset to repositories with at least 10 stars allowed us to focus on projects that received a minimum level of external attention.
Second, we retained only repositories with at least 5 contributors. This constraint was introduced to reduce the presence of purely personal or toy projects and to privilege collaborative development settings, where code quality concerns are more likely to be discussed and addressed explicitly.
Third, we filtered repositories by development activity, retaining only those with a number of commits between 50 and 5000. The lower bound excludes immature or abandoned projects, while the upper bound reduces the impact of anomalous repositories dominated by automated commit activity, and it makes our study scalable.

The repository collection phase was implemented using the GitHub REST API. Since GitHub search queries return at most 1000 results, we adopted a time window slicing strategy based on repository creation date. For periods with a high density of projects, queries were recursively split into smaller temporal windows until the number of matching repositories fell below the API limit. This procedure allowed us to obtain broad coverage of Solidity repositories created between 2018 and 2025.

Starting from the selected repositories, we mined commit histories using \textit{PyDriller}~\cite{spadini2018pydriller}, a framework designed to extract and analyze data from Git repositories. PyDriller provides a structured interface to traverse commit histories and retrieve both metadata and code changes, making it well-suited for large-scale mining of software repository studies.

We retained only commits modifying at least one Solidity source file with extension \texttt{.sol}. The analysis was conducted at commit level, as commits represent a natural unit for capturing localized code evolution together with the intent expressed in commit messages.
To improve the quality of the dataset, we excluded meta commits that do not correspond to atomic development actions. In particular, merge commits were discarded, as they typically aggregate multiple changes and do not reflect a single coherent modification. These commits are automatically identifiable through PyDriller metadata (e.g., multiple parents). Similarly, duplicate commits and commits generated by automated processes were filtered out when identifiable, to avoid introducing noise in the analysis.

For each candidate commit, we extracted relevant metadata, including repository identifier, commit hash, author, branch, and the list of modified files. In addition, we retrieved the source code before and after each modification, enabling a precise comparison of changes. Files used for testing were excluded when identifiable through the \texttt{.t.sol} suffix, since they do not correspond to production smart contracts and may contain optimizations relevant only to local test execution.

\subsection{NLP-based Filtering and Dataset Sampling}

After mining, we applied an NLP-based filtering step to reduce the number of commits to be inspected manually. Since most commits in Solidity repositories are unrelated to gas consumption, semantic filtering was necessary to isolate those more likely to contain optimization-related changes.

We implemented the filtering pipeline using SpaCy and the \texttt{en\_core\_web\_lg} model. Commit messages were processed through lowercasing, tokenization, and lemmatization, so as to normalize lexical variants into a shared base form. Lemmatization was especially important to capture different grammatical realizations of optimization intent, such as \textit{optimize}, \textit{optimized}, and \textit{optimization}.

The filtering rule accepted commits whose lemmatized messages contained the term \textit{gas} and at least one additional lemma associated with optimization intent. The additional vocabulary included terms such as \textit{optimize}, \textit{optimise}, \textit{optimization}, \textit{optimisation}, \textit{cost}, \textit{reduce}, \textit{reduction}, \textit{minimize}, and \textit{opt}. This boolean condition was designed to reduce false positives by excluding commits where gas was mentioned in unrelated contexts.

At the end of this phase, the candidate dataset consisted of 889 commits potentially related to gas optimization. Since a commit may affect multiple Solidity files, the corresponding file level modifications were preserved for subsequent inspection.

Manual inspection of the full refined dataset would still have been too costly and prone to reviewer fatigue. Therefore, we selected a representative sample of 400 commits using proportional stratified sampling. The population of 889 commits was partitioned according to the year of repository creation, and the number of sampled commits for each stratum was computed proportionally to its size in the population. This choice allowed us to preserve the temporal distribution of the original dataset and reduce bias caused by the uneven growth of the Ethereum ecosystem over time.

The selected sample size also provides statistical representativeness of the population. In particular, a sample of 400 instances over a population of 889 ensures a confidence level of 95\% with a margin of error of approximately 5\%, which is commonly considered acceptable in empirical software engineering studies. This supports the reliability of the observed patterns and the generalizability of the findings.

\subsection{Manual Labeling and Validation}
\label{subsec:labeling-validation}

The sampled commits were then analyzed manually to validate their relevance and identify the optimization mechanisms they introduced. For each commit, we first inspected the full commit message and the modified files to confirm that the change was explicitly related to gas optimization.

Within each selected commit, we restricted the analysis to Solidity production files and excluded artifacts not intended for deployment, such as test files, deployment scripts, and configuration files that did not introduce logic changes. We then compared the code before and after the commit in order to identify the specific changes responsible for reducing gas consumption. 
The unit of sampling is the commit, while the unit of manual inspection is the Solidity file modification. Since a single commit may modify multiple files and a single file modification may contain multiple optimization actions, the counts reported in the taxonomy refer to optimization instances rather than commits.
The labeling process involved two independent evaluators. Each evaluator analyzed the commits separately and assigned descriptive labels to the observed optimization actions. A single file modification could involve multiple optimization actions; therefore, the annotation process adopted a multi-label perspective, recording each distinct optimization mechanism independently.

After the independent annotation phase, the evaluators compared their labels and resolved discrepancies through discussion. This consensus-based process ensured a shared interpretation of the observed changes and reduced subjectivity in the classification. For every analyzed file, we documented whether it was a true positive, described the relevant code changes in natural language, and associated each change with a preliminary technical label.

\subsection{Open Coding and Taxonomy Construction}

To answer \textbf{RQ$_1$}, we analyzed the manually annotated changes through an open coding procedure. Instead of starting from predefined optimization categories, we derived labels directly from the observed code transformations. Similar labels were iteratively grouped into higher-level categories based on technical affinity.
The construction of the taxonomy was also carried out collaboratively. The two evaluators jointly reviewed the preliminary labels and organized them into macro categories and subcategories through iterative discussion. This process allowed the taxonomy to emerge from the data while ensuring consistency and agreement in the categorization.

This bottom up approach resulted in an empirical taxonomy of gas optimization strategies grounded in real development practice. The taxonomy captures both local code level interventions and broader design level changes, reflecting the heterogeneous nature of gas optimization in Solidity smart contracts.

At the end of this step, the final output of the study is a structured classification of the recurring strategies used by developers to reduce execution costs in Ethereum smart contracts.

\section{Experimental Results}
\label{sec:results}
In this section, we report the results of our work delivering the taxonomy we produced alongside a comprehensive explanation of each category and subcategory.

\begin{figure*}[htb!]
    \centering
    \includegraphics[width=\textwidth]{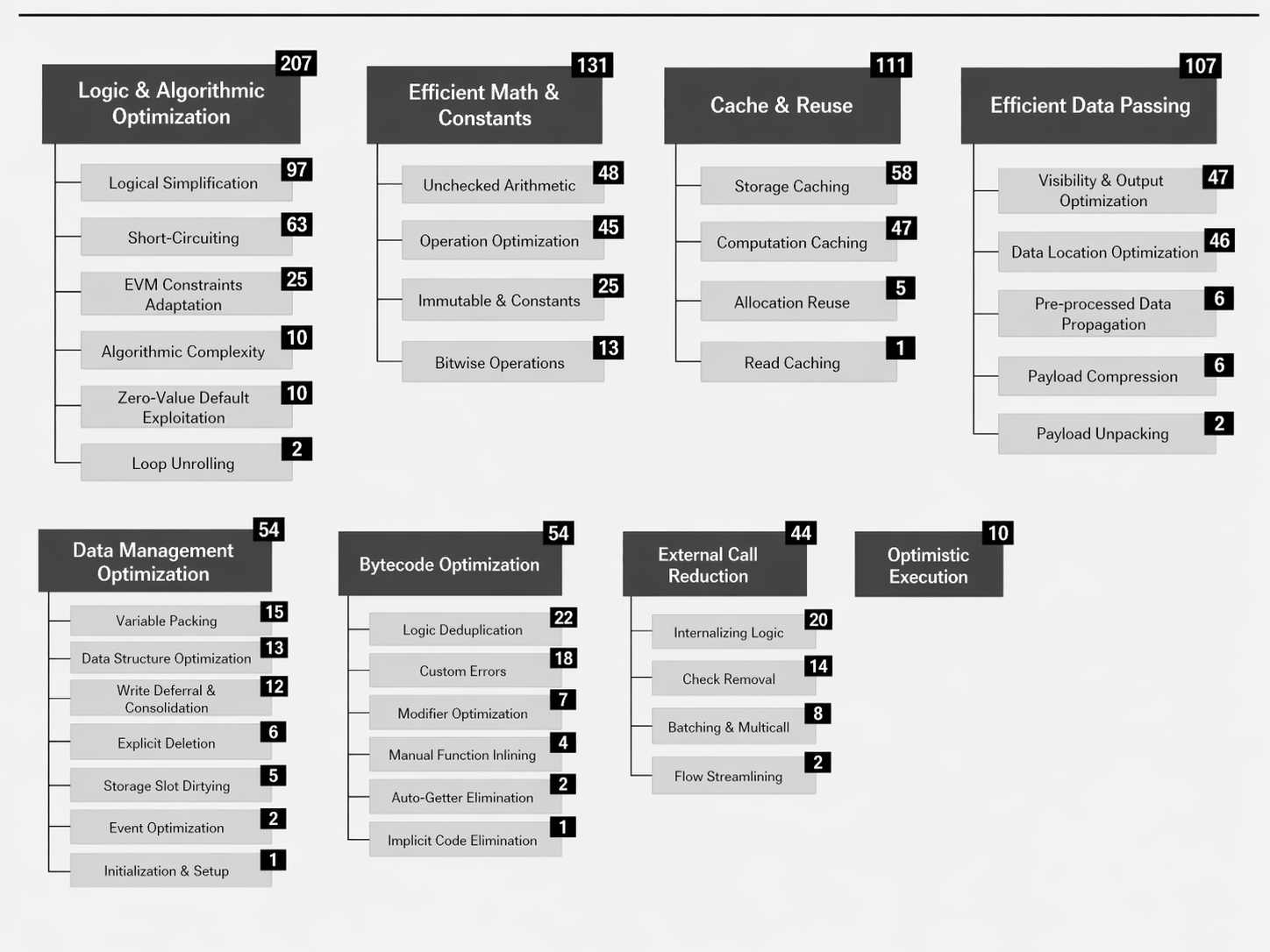}
    \caption{Empirical taxonomy of gas optimization strategies derived from commit-level analysis.}
    \label{fig:taxonomy}
\end{figure*}

The analysis of the sampled commits resulted in an empirical taxonomy of gas optimization strategies observed in Solidity smart contracts. Each optimization action identified during manual inspection was assigned to a category and, when applicable, to a more fine-grained subcategory. Since multiple optimizations can occur within the same file, the counts reported in this section refer to the total number of observed optimization instances rather than the number of commits.
The taxonomy is organized into macro categories reflecting different levels of intervention, ranging from low-level arithmetic optimizations to architectural changes.
Figure~\ref{fig:taxonomy} presents a visual representation of the resulting taxonomy, illustrating the hierarchical organization of categories and subcategories.

In the following, we describe each category together with its corresponding subcategories, and we report the number of file modification that falls into each category. Clearly, the gas optimizations enclosed in the modification of one commit can fall into diverse categories of the created taxonomy.

\subsection{Logic and Algorithmic Optimization (207)}

This category captures optimization strategies that act directly on the execution flow and computational logic of smart contracts. Unlike other categories that focus on data handling or storage costs, these optimizations aim at reducing the number of operations executed by the Ethereum Virtual Machine, simplifying control structures, and eliminating redundant or unnecessary computation steps. With 207 optimization instances, this emerges as the most prominent category, highlighting the central role of control flow design in achieving gas efficiency. 
\\\textbf{Logical Simplification (97).} Transformations simplify the logical structure of the code without altering its functionality by removing redundant operations, unnecessary assignments, and superfluous conditional branches, resulting in a more direct execution path. Typical examples include eliminating no-op statements, collapsing nested conditions, and removing computations whose results are not used; although each change yields limited savings, their frequency indicates a consistent effort to reduce computational overhead at the micro level. 
\\\textbf{Short-Circuiting (63).} Conditional checks are introduced or reordered to terminate execution as early as possible when further computation is unnecessary, following a fail-fast principle where inexpensive conditions are evaluated before more expensive operations such as storage access or external calls, preventing the execution of costly operations in unfavorable scenarios and reducing gas consumption in edge cases and invalid inputs.
\\\textbf{EVM Constraints Adaptation (25).} Structural adaptations are introduced to comply with constraints imposed by the Ethereum Virtual Machine, preventing execution failures or deployment limitations that would otherwise result in wasted gas; these include splitting computations across multiple steps or transactions, restructuring loops to avoid gas limits, and adjusting execution patterns to fit within architectural constraints. 
\\\textbf{Algorithmic Complexity Reduction (10).} Changes target the asymptotic complexity of algorithms by redesigning the underlying logic to reduce the total number of operations required; typical transformations include replacing iterative searches with constant-time access patterns, reducing nested loops, and restructuring data access for more efficient scaling, yielding significant savings when applied to large datasets or frequently executed functions.
\\\textbf{Zero-Value Default Exploitation (10).} Optimizations exploit default values implicitly provided by the EVM, such as zero-initialized storage or memory, removing explicit initialization or redundant checks when the same behavior is already guaranteed by default semantics and avoiding unnecessary instructions. \\\textbf{Loop Unrolling (2).} Iterative constructs are replaced with explicitly expanded sequences of operations, eliminating loop control overhead such as index updates and boundary checks at the cost of increased code size; although rarely observed, this technique can be effective in small fixed-size iterations where minimizing runtime overhead is critical. A substantial portion of gas optimization effort is therefore devoted to improving the logical structure of smart contracts, with developers achieving savings by restructuring control flow and eliminating unnecessary computation.

\subsection{Efficient Math and Constants (131)}

This category captures optimizations that reduce gas consumption by improving the efficiency of arithmetic operations and the handling of constant values. Rather than modifying the overall logic of the program, these techniques focus on replacing standard but costly operations with more efficient alternatives or avoiding unnecessary access to expensive storage locations. With 131 instances, this category highlights the importance of low-level arithmetic awareness in gas optimization. \\\textbf{Unchecked Arithmetic (48).} Arithmetic operations are wrapped in \texttt{unchecked} blocks to bypass overflow and underflow checks automatically introduced by the Solidity compiler since version 0.8.0; when developers can guarantee that operations remain within safe bounds, removing these checks reduces execution overhead, especially in loops and bounded computations where correctness is ensured by design. \\\textbf{Operation Optimization (45).} Standard arithmetic expressions are replaced with more efficient equivalents by simplifying formulas, removing redundant computations, and restructuring calculations to minimize the number of executed operations; these changes preserve the overall complexity while reducing the cost of individual instructions, leading to consistent micro-level savings. \\\textbf{Immutable and Constants (25).} Values that do not change are declared as \texttt{constant} or \texttt{immutable}, allowing them to be embedded directly into the contract bytecode or initialized at deployment time, thereby avoiding expensive \texttt{SLOAD} operations and enabling cheaper stack-level access when the same value is used repeatedly. \\\textbf{Bitwise Operations (13).} Bitwise operators are used to replace more expensive arithmetic operations, with shift operations acting as efficient alternatives to multiplication or division by powers of two and bitwise logic enabling compact management of flags and packed data structures; although less frequent, these optimizations reflect a deeper understanding of EVM execution costs and low-level computation. The distribution of these patterns shows that arithmetic optimization is often achieved by trading safety checks and abstraction for performance when correctness can be statically guaranteed.

\subsection{Cache and Reuse (111)}

This category captures strategies aimed at avoiding redundant computations and repeated access to expensive data sources. The underlying principle is to compute or retrieve a value once, store it in a temporary variable, and reuse it whenever needed. With 111 instances, this category confirms that reducing repetition is a key optimization strategy in Solidity development.
\\\textbf{Storage Caching (58).} Values stored in persistent storage are read once and cached in a local variable, avoiding repeated \texttt{SLOAD} operations, which are among the most expensive in the EVM; typical examples include caching state variables or array elements before using them multiple times within a function. \\\textbf{Computation Caching (47).} Results of computations such as arithmetic expressions, hash calculations, or function calls are stored and reused instead of being recomputed multiple times, eliminating redundant execution and reducing the overall computational cost. 
\\\textbf{Allocation Reuse (5).} Already allocated memory is reused instead of creating new variables or data structures, reducing allocation overhead and improving execution efficiency in functions with complex data handling.
\\\textbf{Read Caching (1).} Repeated read operations from non-storage sources, such as memory or calldata, are cached to avoid repeated indexing and access overhead, providing marginal but consistent savings. A significant portion of gas optimization effort is therefore dedicated to eliminating redundancy, with developers achieving substantial savings by minimizing repeated storage access and avoiding recomputation without altering the overall program logic.

\subsection{Efficient Data Passing (107)}

This category captures optimizations related to how data is transferred between functions and contracts. In Ethereum, data movement is inherently costly, both in terms of transaction payload size and memory handling during execution. These techniques aim to minimize data copying, reduce payload size, and streamline the interface between caller and callee. With 107 instances, this category highlights the importance of data flow design in gas-efficient smart contract development. \\\textbf{Data Location Optimization (46).} Optimizations leverage the different cost models of Solidity data locations, with a preference for \texttt{calldata} over \texttt{memory} for read-only inputs, avoiding unnecessary memory allocation and copying and reducing both execution cost and memory overhead. \\\textbf{Visibility and Output Optimization (47).} Function interfaces are refined to reduce communication overhead, using \texttt{external} visibility to avoid implicit memory copies and simplifying or removing return values to minimize encoding costs. \\\textbf{Pre-processed Data Propagation (6).} Data is pre-processed before being passed to a function or contract, shifting computation upstream and reducing the amount of processing required by the callee. \\\textbf{Payload Compression (6).} Multiple values are compacted into smaller representations or more efficient layouts to reduce the number of bytes transmitted, lowering transaction costs since data size directly impacts gas consumption. \\\textbf{Payload Unpacking (2).} Decoding logic is shifted outside the contract, allowing the contract to receive already unpacked inputs and avoid costly parsing operations, reflecting a tradeoff between data transmission and computation. These patterns show that developers actively optimize the boundary between contracts and their environment, reducing both data transfer costs and internal processing overhead.

\subsection{Data Management Optimization (54)}

This category focuses on optimizing how data is stored, structured, and updated within the contract. Since storage operations are among the most expensive in the EVM, efficient data management plays a crucial role in reducing gas consumption. Although less frequent than other categories, these optimizations often provide significant savings due to the high cost of storage access. \\\textbf{Variable Packing (15).} Multiple variables are packed into the same storage slot by carefully ordering them and selecting appropriate data types, reducing the number of storage slots used and lowering both read and write costs. \\\textbf{Data Structure Optimization (13).} Data structures are redesigned to better match access patterns, replacing inefficient constructs with alternatives such as mappings for constant-time access, thereby reducing both computational overhead and storage usage. \\\textbf{Write Deferral and Consolidation (12).} Storage writes are minimized by performing intermediate computations locally and committing results to storage only once, significantly reducing costs associated with repeated state updates. \\\textbf{Explicit Deletion (6).} Storage variables are explicitly cleared when no longer needed, enabling gas refunds and reducing long-term storage footprint while maintaining a cleaner on-chain state. \\\textbf{Storage Slot Dirtying (5).} Optimizations exploit the EVM cost model where repeated writes to the same storage slot within a transaction are cheaper after the first modification, grouping updates to reduce overall cost. \\\textbf{Event Optimization (2).} Event emission is optimized by reducing logged data or adjusting indexing strategies to avoid unnecessary overhead, limiting the cost of logging operations. \\\textbf{Initialization and Setup (1).} Contract initialization is streamlined by reducing unnecessary assignments and structuring the initial state more efficiently, lowering deployment or setup costs. These findings confirm that efficient data management is a critical component of gas optimization, with developers achieving substantial savings by reducing storage footprint and minimizing write operations in state-intensive contracts.

\subsection{Bytecode Optimization (54)}

This category captures optimizations that target the structure and size of the generated bytecode. In Ethereum, the cost of deploying a smart contract is directly proportional to its bytecode size, and contracts are also subject to strict size limits. These optimizations aim to reduce code duplication and represent functionality in a more compact form. With 54 instances, this category highlights that developers consider not only runtime efficiency but also deployment overhead. \\\textbf{Logic Deduplication (22).} Duplicated logic is eliminated by refactoring repeated code into shared internal functions or reusable components, reducing the overall bytecode footprint; although this may introduce a minor runtime overhead due to additional jumps, it significantly lowers deployment cost by avoiding redundant instruction sequences. \\\textbf{Custom Errors (18).} Revert strings are replaced with custom errors, allowing developers to define compact identifiers instead of storing verbose messages in the bytecode, reducing contract size and improving efficiency, especially in contracts with numerous validation checks. \\\textbf{Modifier Optimization (7).} Modifier logic is restructured to reduce duplication caused by inlining into multiple functions, often by delegating checks to internal functions, preserving readability while minimizing bytecode growth. \\\textbf{Manual Function Inlining (4).} Small functions are manually inlined to eliminate function call overhead, embedding logic directly into the caller; while this increases code size, it can improve runtime efficiency in performance-critical paths. \\\textbf{Auto-Getter Elimination (2).} Automatically generated getter functions for public state variables are removed when not needed, reducing unnecessary bytecode and lowering deployment cost. \\\textbf{Implicit Code Elimination (1).} Unused or implicitly generated code is removed to further compact the contract, contributing marginally to reducing deployment overhead. These patterns indicate that developers actively manage code size and structure, applying refactoring techniques to produce more compact and efficient bytecode.

\subsection{External Call Reduction (44)}

This category includes optimizations aimed at reducing interactions with external contracts. External calls are expensive because they require the creation of a new execution context and involve additional overhead for context switching and security checks. Although less frequent than other categories, these optimizations often yield significant savings due to the high cost of cross-contract interactions. \\\textbf{Internalizing Logic (20).} Logic previously delegated to external contracts is moved into the current contract, eliminating cross-contract communication and reducing execution overhead, particularly when the external functionality is simple enough to be executed locally. \\\textbf{Check Removal (14).} Redundant validation calls to external contracts are removed when the subsequent operation would fail anyway if conditions are not met, avoiding unnecessary calls and reducing verification costs. \\\textbf{Batching and Multicall (8).} Multiple external interactions are grouped into a single call using batching mechanisms or multicall patterns, reducing the cumulative overhead associated with repeated context switching. \\\textbf{Flow Streamlining (2).} Interaction flows between contracts are simplified by reducing intermediate steps or unnecessary calls, leading to more direct and efficient execution paths. These findings show that developers actively minimize external interactions by restructuring contract architecture and eliminating redundant calls, with even small reductions in cross-contract communication resulting in substantial gas savings.

\subsection{Optimistic Execution (10)}

This category captures optimizations where developers remove explicit checks or validations to reduce the cost of the common execution path. Instead of verifying conditions beforehand, the contract directly performs the operation and relies on the EVM to revert in case of failure.
Although relatively infrequent, this pattern reflects a deliberate tradeoff between safety checks and efficiency, favoring lower cost in the typical execution scenario.

\section{Discussion}
\label{sec:discussion}
This section discusses the main implications of our findings, focusing on how developers optimize gas in practice and on the trade-offs between cost, correctness, maintainability, and execution context. 

\subsection{Inline Assembly as a Precision Tool}

Inline assembly, through \texttt{yul} blocks, represents the lowest-level mechanism available to Solidity developers for gas optimization. By operating directly on the Ethereum Virtual Machine, developers bypass compiler abstractions and interact with opcodes, gaining fine-grained control over execution.

From a theoretical perspective, this approach enables full control over storage layout, memory allocation, and external calls. However, our empirical observations suggest that its use in practice is highly limited and targeted. Inline assembly appears only in a small fraction of the analyzed files, indicating that developers rarely rely on it as a general optimization strategy.

Instead, assembly is primarily used in specific contexts where its benefits clearly outweigh its risks. In particular, it is concentrated in arithmetic operations and low level logic. These scenarios operate mostly on the stack, which is the cheapest execution context in the EVM, and allow developers to exploit direct opcode manipulation without introducing excessive complexity.

Three recurring motivations for using assembly emerge from the data. First, developers use it to bypass redundant safety checks introduced by the Solidity compiler, such as overflow and underflow protections. When correctness can be guaranteed by design, removing these checks reduces execution cost. Second, assembly enables efficient bitwise operations, allowing compact manipulation of flags and packed variables. Third, it provides direct access to calldata, avoiding the overhead of ABI decoding.

Despite these advantages, assembly is rarely used beyond localized interventions. Its adoption at architectural level would significantly increase the likelihood of critical errors, such as incorrect memory offsets or unintended overwrites of storage. Moreover, assembly introduces substantial maintenance challenges, as it reduces code readability and complicates auditing processes. These trade-offs explain why developers restrict their usage to well-defined and performance-critical segments of the code.

\subsection{Context Dependent Optimization Strategies}

Gas optimization does not follow a fixed set of universally applicable rules. Instead, its effectiveness depends strongly on the execution context and on the trade off between the frequency of an operation and its cost. Our analysis reveals that developers often adopt contrasting strategies depending on this balance.

A first strategy can be described as \textit{optimistic execution}. In this case, developers remove conditional checks when the probability of failure is low and the cost of the operation is limited. The rationale is to reduce the baseline cost of frequently executed successful transactions. Even small savings per execution can accumulate into significant gains when operations are invoked repeatedly. The trade off is that, in the rare case of failure, the transaction may revert without providing detailed feedback.
A second, complementary strategy corresponds to \textit{short circuiting}. Here, developers introduce explicit checks to prevent the execution of expensive operations when certain conditions are not met. This pattern is particularly relevant when subsequent operations involve costly actions such as storage writes or external calls. By paying a small upfront cost for validation, developers avoid much larger expenses in unfavorable scenarios.

These two strategies reflect a shared underlying principle: optimization decisions are driven by the balance between execution frequency and computational cost. Rather than applying fixed rules, developers reason about expected usage patterns and tailor their code accordingly.

\subsection{Implications for Gas Optimization Practices}

The findings of this study highlight that gas optimization is inherently context-dependent and spans multiple levels of abstraction. Low-level techniques, such as inline assembly, provide strong performance benefits but are applied selectively due to their risks. At the same time, higher-level decisions, such as the placement or removal of conditional checks, demonstrate that even simple code transformations can have a significant economic impact when guided by execution patterns.
This suggests that effective optimization cannot be reduced to isolated micro improvements or static recommendations. Instead, it requires an understanding of how contracts are used in practice, how frequently specific paths are executed, and how different operations contribute to overall cost.

From a tooling perspective, these observations indicate that current approaches based solely on detecting inefficient patterns may be insufficient. Future tools should incorporate contextual information and support developers in reasoning about trade-offs between cost, frequency, and correctness, rather than focusing exclusively on local code transformations.

\section{Threats to Validity}
\label{sec:treats}

In this section, we delineate the principal threats that may compromise the validity of our study and describe the mitigation strategies implemented throughout the phases of data collection, manual annotation, and taxonomy development.
\\\textbf{Construct Validity.}
Threats to construct validity mainly concern the way concepts were identified, interpreted, and organized into taxonomy categories. Since the taxonomy was derived through an open coding process, the identification of relevant concepts from the analyzed sources may be influenced by the subjective interpretation of the evaluators. To mitigate this threat, two evaluators independently analyzed the selected sources and identified the concepts to be included in the taxonomy. Disagreements were discussed until a consensus was reached. In addition, the taxonomy was iteratively refined during the analysis to ensure that concepts were consistently interpreted and organized across the dataset.

Another possible threat relates to the level of abstraction used to represent the identified concepts. Some concepts may have been described at a more specific or more general level depending on the wording used in the analyzed sources. To reduce this issue, similar concepts were compared and merged only after discussion among the authors. Moreover, each final taxonomy category was traced back to the original evidence, ensuring that the resulting taxonomy remained grounded in the analyzed data.
\\\textbf{Internal Validity.}
Internal validity threats concern the correctness and consistency of the process used to build the taxonomy. A possible threat is that some relevant concepts may have been overlooked during the coding phase, especially when they appeared only implicitly in the analyzed material.

Another threat is related to researcher bias during the aggregation of codes into higher-level categories. Since this step requires interpretation, different researchers might organize the same codes differently. To address this issue, the grouping of concepts was discussed among the authors, and conflicts were resolved collaboratively. The taxonomy was revised until the authors agreed that the categories were coherent, non-overlapping where possible, and representative of the underlying data.
\\\textbf{External Validity.}
Threats to external validity concern the generalizability of the proposed taxonomy. The taxonomy was derived from the selected set of sources and may therefore reflect the characteristics, terminology, and practices reported in those sources. As a result, it may not fully capture concepts that appear in other domains, software ecosystems, industrial contexts, or more recent studies not included in our dataset.

Another possible limitation is that the taxonomy reflects the current state of the analyzed literature and artifacts. Since software engineering practices evolve over time, new categories may emerge in future work. Therefore, the taxonomy should be interpreted as an extensible classification rather than a definitive and immutable one.

Moreover, the use of open coding supports the discovery of concepts grounded in the analyzed data, but the resulting taxonomy may still depend on the specific selection criteria adopted in the study. To mitigate this threat, we clearly defined the inclusion and exclusion criteria and documented the process used to select the analyzed sources. Nevertheless, future replications on additional datasets or in different software engineering contexts may lead to refinements or extensions of the proposed taxonomy.

\section{Conclusion}

\label{sec:conclusion}
This paper presented an empirical study on gas optimization practices in Solidity smart contracts, based on the analysis of commit-level changes extracted from open source repositories. By combining repository mining, NLP-based filtering, stratified sampling, and manual inspection with open coding, we derived a taxonomy of optimization strategies grounded in real development activity composed of 8 macro categories and 36 subcategories.

The results show that gas optimization is not limited to low-level micro-optimizations but spans multiple layers of the software stack. A large portion of optimizations focuses on control flow and algorithmic improvements, indicating that developers frequently reduce costs by simplifying execution paths and eliminating unnecessary computation. At the same time, recurring patterns such as caching, efficient data passing, and arithmetic tuning confirm that developers actively reason about the cost model of the Ethereum Virtual Machine when writing and revising code.
Another key finding is that optimization decisions are strongly influenced by context. Developers balance execution frequency and operation cost when deciding whether to introduce or remove checks, restructure logic, or reduce external interactions. This highlights that gas optimization cannot be captured entirely through static rules or isolated patterns, but requires an understanding of how contracts are used in practice.

The study also shows that certain powerful techniques, such as inline assembly, are applied selectively rather than broadly. While they enable fine-grained control and potential savings, their complexity and associated risks limit their use to well-defined scenarios. This reinforces the idea that developers favor maintainability and safety, applying aggressive optimizations only where the benefits are clear.
The taxonomy produced in this work provides a structured view of gas optimization strategies observed in practice and can serve as a reference for both researchers and practitioners. For researchers, it offers an empirical basis for future studies on smart contract optimization and tool support. For practitioners, it provides a consolidated overview of techniques that can be applied to improve gas efficiency.

Future work may extend this study by investigating the effectiveness of individual optimization strategies in terms of quantitative gas savings or by integrating contextual information into automated tools to better support developer decision-making. Another relevant direction concerns the analysis of correctness and reliability trade-offs, with particular attention to whether certain optimizations may unintentionally introduce bugs or security weaknesses. Finally, the taxonomy derived in this study could be used as the foundation for the development of automated gas optimization tools capable of combining multiple optimization strategies while adapting them to different execution contexts.

\section*{Declaration on Generative AI}
\label{sec:declaration-generative-ai}

During the preparation of this work, the authors used generative AI tools only for grammar and spelling checks. After using these tools, the authors reviewed and edited the content as needed and take full responsibility for the publication's content.

\section*{Acknowledgment}
    This project was partially funded through TruMaN (Italian Ministry of University and Research, 2022, PRIN, Project 2022F5CLN2).

\bibliography{9.bibliography}

\end{document}